\documentclass{optica-article}

\journal{opticajournal} 

\articletype{Research Article}

\usepackage{setspace}
\usepackage{lineno}
\usepackage{url}
\usepackage{makecell}
\usepackage{graphicx}
\usepackage{epstopdf}
\usepackage{float}
\usepackage{stfloats}

\usepackage{hyperref}

\usepackage{amsmath,amsfonts}
\usepackage{mathrsfs}
\usepackage{bm}

\usepackage{subcaption}
\usepackage{tikz}
\usepackage{booktabs}
\usepackage{multirow}
\usepackage[table,xcdraw]{xcolor} 
\usepackage{anyfontsize}
\usepackage{algorithm}
\usepackage{algorithmicx}
\usepackage{algpseudocode}

\usepackage{listings}
\usepackage{textcomp}
\usepackage{manyfoot}
\usepackage{pifont}
\usepackage{siunitx}
\usepackage{etoolbox}
\begin{document}
\title{Measurement-Adaptive Light-Cone Transform for Robust Non-Line-of-Sight Imaging}
\author{
Lianfang Wang\authormark{1,$\dagger$}, Xueying Liu\authormark{1,$\dagger$}, Junnan Dong\authormark{1},
Qinghua Zhang\authormark{1},
Jun Liu\authormark{1},  Yong Wang\authormark{2,*}, and Yuping Duan\authormark{1,*}}
\address{
\authormark{1}School of Mathematical Sciences, Beijing Normal University, Beijing 100875, China\\
\authormark{2}School of Physics, Nankai University, Tianjin 300071, China\\
\authormark{$\dagger$}These authors contributed equally to this work}
\email{\authormark{*}Corresponding author: yongwang@nankai.edu.cn; doveduan@gmail.com} 

\begin{abstract*} 
Practical non-line-of-sight (NLOS) measurements are affected
by noise and deviations from idealized light transport,
which can degrade reconstruction with conventional
light-cone transform (LCT) methods. We propose the
Measurement-Adaptive Light-Cone Transform (MALCT), a
physics-guided framework that retains the efficient
Fourier-domain inversion structure of LCT. 
{{Within a
unified variational formulation, MALCT combines the
analytical response with a shared learned response
through measurement-conditioned weighting and predicts
a measurement-conditioned, frequency-dependent
regularization map.}}
An attention-enhanced spatio-temporal feature extraction module captures long-range temporal dependencies and local spatial correlations to support reconstruction under low-signal conditions. Experiments
on synthetic data and real measurements from different
NLOS systems demonstrate improved reconstruction quality.
Controlled evaluations further assess robustness to
temporal-response broadening and reduced acquisition
time. MALCT also preserves scene structures with spatial
sampling grids down to $32\times32$, supporting its
potential for NLOS imaging under limited acquisition
budgets.
\end{abstract*}

\section{Introduction}
Non-line-of-sight (NLOS) imaging reconstructs hidden scenes
from indirectly reflected light. A visible relay surface,
such as a wall, redirects illumination toward an occluded
scene and returns part of the scattered light to the detector.
These indirect measurements encode information about the
geometry and reflectivity of hidden objects. The ability
to recover this information has motivated research in
applications such as autonomous navigation,
remote sensing, and biomedical imaging ~\cite{bauer2015non,shi2025fast,lindell2019acoustic,
chan2017non,maeda2019recent}.

NLOS systems can be classified by their illumination and
acquisition configurations. Passive approaches exploit
ambient illumination or naturally occurring intensity
variations ~\cite{batarseh2018passive,boger2019passive,
beckus2019multi,wang2026passive}, whereas active approaches
use controlled illumination. Active transient systems
typically combine pulsed illumination with synchronized,
time-resolved detection to record the spatio-temporal
distribution of returning light 
~\cite{laurenzis2014feature,
ou2022computational,wang2024enhancing,sun2026transient}.
These systems are further categorized as confocal or
non-confocal according to whether the illumination and
detection positions coincide on the relay surface.
The recorded transients provide time-of-flight information
for reconstructing hidden-scene geometry and reflectivity.

Recovering a hidden scene from transient measurements is
an ill-posed inverse problem. The light-cone transform
(LCT) exploits a convolutional representation of confocal
light transport after spatio-temporal resampling, enabling
efficient Fourier-domain reconstruction 
~\cite{o2018confocal}. Subsequent developments extend
or refine this modeling framework 
~\cite{young2020non,sultan2024towards}. Wave-based and
phasor-field methods instead formulate reconstruction
through wave propagation and diffraction-inspired models 
~\cite{lindell2019wave,liu2020phasor,
luesia2022non,jiang2021ring}. Although these approaches
provide efficient reconstruction, their performance can
deteriorate when measurements are noisy, sparsely sampled,
or inconsistent with the assumed forward model.

To improve reconstruction under limited measurements,
prior-based approaches incorporate sparsity, smoothness,
geometric constraints, or statistical models 
~\cite{
ye2021compressed,pei2021dynamic,liu2022photon,
liu2023few,ding2024curvature}. Iterative implementations
often require repeated forward and adjoint evaluations
and careful regularization selection. Learning-based
methods learn mappings from transient measurements to
depth, reflectivity, or volumetric representations 
~\cite{grau2020deep,zhu2023compressive}, while
physics-guided methods combine analytical light-transport
models with trainable components%
~\cite{shen2021non,yu2023enhancing,su2024model,
cui2025transdiff,gao2026learned}. These developments
motivate reconstruction strategies that retain the
computational structure of analytical inversion while
adapting to measurement quality. Recovering weak returns,
depth discontinuities, and fine structures remains
challenging, particularly under reduced acquisition times,
sparse spatial sampling, and changes in acquisition systems.

\begin{figure}[t]
\centering
\includegraphics[width=\linewidth]{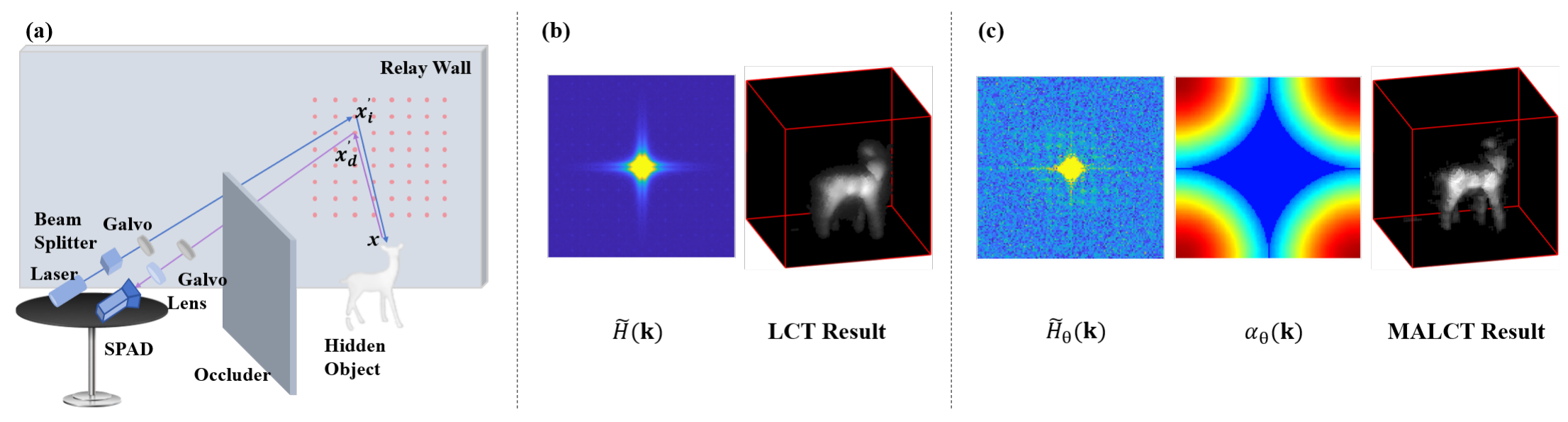}
\caption{
{Overview of the proposed framework.
(a) Confocal NLOS imaging configuration.
(b) Conventional LCT reconstruction with a fixed analytical
Fourier response $\widetilde{H}$ and scalar regularization.
(c) MALCT reconstruction with a measurement-conditioned
effective response $\widetilde{H}_{\Theta}(\mathbf{k};\tau)$
and regularization map $\alpha_{\Theta}(\mathbf{k};\tau)$.
The effective response combines the analytical response
with a shared learned response through
measurement-conditioned weighting.}
}
\label{nlos-figure}
\end{figure}

Conventional LCT uses a fixed analytical Fourier response
and a scalar regularization parameter. This formulation
enables efficient pointwise inversion, but offers limited
flexibility when practical measurements deviate from the
assumed model or exhibit different noise characteristics.
Such deviations can arise from non-ideal reflection,
geometric mismatch, temporal calibration errors, and
detector response. Meanwhile, the reliability of spectral
components can vary across frequencies and measurements.
A single regularization strength may therefore suppress
weak structural information at some frequencies while
providing insufficient noise suppression at others.

In this work, we propose the Measurement-Adaptive Light-Cone Transform (MALCT), a physics-guided extension of LCT~\cite{o2018confocal}. As illustrated in Figure~\ref{nlos-figure}, MALCT retains the efficient Fourier-domain inversion structure within a unified variational formulation. It combines the analytical response with a shared learned response through measurement-conditioned weighting and predicts a positive regularization map conditioned on both frequency and the input transient. These components jointly adapt the effective inverse filter to each measurement. The learned response provides an effective representation for reconstruction, rather than an
explicit estimate of individual physical mismatch sources. An attention-enhanced spatio-temporal feature extraction module captures long-range temporal dependencies and local spatial correlations to guide this adaptation. The network is trained exclusively on synthetic data and evaluated on unseen synthetic scenes and real measurements from different NLOS systems. Controlled experiments assess reconstruction robustness to Gaussian IRF broadening and reduced acquisition time, while
ablation studies examine the contributions of attention, the learned response, and regularization.

To sum up, the main contributions are as follows:
\begin{itemize}

\item
We formulate an adaptive LCT inversion that combines measurement-conditioned response weighting with a  measurement-conditioned frequency-dependent regularization map, while preserving the diagonal Fourier-domain structure.

\item
We develop a physics-guided framework that integrates attention-enhanced spatio-temporal feature extraction, adaptive inversion, and volume projection for joint intensity and depth reconstruction.

\item
We evaluate the framework on synthetic and real measurements, including controlled IRF and acquisition-time experiments and spatial sampling grids down to $32\times32$. Component ablations further assess the roles of attention, response learning, and regularization.

\end{itemize}

\section{Preliminaries}
The forward model of NLOS transient imaging describes
light transport between a visible relay surface and a
hidden scene. Let $\bm{x}=(x,y,z)\in\Omega\subset\mathbb{R}^3$
denote a point in the hidden-scene domain. An ultrashort
laser pulse illuminates a relay-wall point
$\bm{x}_i'=(x_i',y_i',0)$, and the returning light is
recorded at $\bm{x}_d'=(x_d',y_d',0)$ on the same planar
wall. The confocal case is illustrated in
Figure~\ref{nlos-figure}(a).

Under Lambertian reflection, a single reflection at the
hidden scene, negligible occlusion, and ideal temporal
detection, a simplified noise-free transient model is
\begin{equation}
\tau_0(\bm{x}_i',\bm{x}_d',t)
=\iiint_\Omega\frac{
\langle\bm{x}_i'-\bm{x},\bm{n}(\bm{x})\rangle_{+}
}{
\|\bm{x}_i'-\bm{x}\|_2^3
}
\frac{
\langle\bm{x}_d'-\bm{x},\bm{n}(\bm{x})\rangle_{+}
}{
\|\bm{x}_d'-\bm{x}\|_2^3
}
\delta\!\left(
\frac{
\|\bm{x}-\bm{x}_i'\|_2+
\|\bm{x}-\bm{x}_d'\|_2
}{c}-t
\right)
u(\bm{x})\,\mathrm{d}\bm{x},
\label{nonconfocal-FM}
\end{equation}
where $u(\bm{x})$ denotes an effective reflectivity
density supported on the hidden scene,
$\bm{n}(\bm{x})$ is the unit surface normal,
$c$ is the speed of light, and $\delta(\cdot)$ enforces
the time-of-flight constraint. The notation
$\langle\bm{a},\bm{b}\rangle_+
:=\max\{\langle\bm{a},\bm{b}\rangle,0\}$
restricts the angular factors to the reflecting
hemisphere. The time coordinate is referenced to the
relay wall, with the known source-to-wall and
wall-to-detector delays removed. Constant radiometric
factors are omitted, and relay-wall angular effects
are assumed to be compensated or approximated as constant.

Equation~\eqref{nonconfocal-FM} allows distinct illumination
and detection positions. In the confocal configuration,
these positions coincide,
$\bm{x}'=\bm{x}_i'=\bm{x}_d'$.
To obtain the conventional LCT model, we adopt an
isotropic approximation that neglects the directional
dependence of the hidden-surface reflection.
The resulting model is
\begin{equation}
\tau_0(\bm{x}',t)
=
\iiint_\Omega
\frac{u(\bm{x})}{\|\bm{x}'-\bm{x}\|_2^4}
\delta\!\left(
\frac{2\|\bm{x}'-\bm{x}\|_2}{c}-t
\right)
\,\mathrm{d}\bm{x},
\label{eqconnlos}
\end{equation}
where $u$ now denotes the effective reflectivity under
this approximation.

Following the LCT formulation of
O'Toole et al.~\cite{o2018confocal}, appropriate amplitude
weighting and quadratic resampling of the temporal and
depth coordinates convert this model into a
three-dimensional shift-invariant convolution.
The corresponding discrete measurement model on the
original detector-time grid is
\begin{equation}
\tau=R_t^{-1}HR_z u+\eta,
\label{forward-model}
\end{equation}
where $\tau$ denotes the measured transient data,
$H$ is the convolution operator in the resampled domain,
and $\eta$ collects deviations from the nominal model,
including measurement noise and unmodeled acquisition
effects. The operators $R_t$ and $R_z$ incorporate the
temporal and depth resampling, respectively, together
with the associated LCT weights. The notation
$R_t^{-1}$ denotes the corresponding mapping back to
the original detector-time grid; in a discrete
implementation, it need not be an exact matrix inverse.

\section{Our method}\label{method}

\subsection{Frequency-domain variational formulation}
Recovering the hidden reflectivity volume $u$ from the
transient measurement $\tau$ in Eq.~\eqref{forward-model}
is an ill-posed inverse problem. We introduce the
resampled variables
$u_r=R_z u$, $\tau_r=R_t\tau$, and $\eta_r=R_t\eta$.
For notational simplicity, the subscript $r$ is omitted
throughout this subsection. Accordingly, $u$, $\tau$,
and $\eta$ denote quantities in the resampled LCT
domain. We use $\mathbf{x}=(x,y,s)$ to denote coordinates
on this common computational grid, where $s$ represents
the resampled axial coordinate, and
$\mathbf{k}=(k_x,k_y,k_s)$ denotes the corresponding
Fourier-frequency coordinates. The nominal forward
model is
\begin{equation}
\tau(\mathbf{x})
=
(Hu)(\mathbf{x})+\eta(\mathbf{x}),
\label{kernel}
\end{equation}
where $H$ is a three-dimensional convolution operator.
The temporal and depth resampling place the measurement
and reflectivity volume on compatible grids for this
convolutional representation.

We adopt an unweighted quadratic data-fidelity term
as a tractable measure of the discrepancy between the
predicted and observed transients. Although this term
can be derived from a zero-mean isotropic Gaussian
residual model, we use it here as a surrogate rather
than an exact statistical description of the measurements.
In particular, the synthetic degradation model includes
a background offset and signal-dependent perturbations,
while temporal resampling can introduce correlations
between noise components. These effects are not fully
represented by an unweighted quadratic penalty. Applying the three-dimensional Fourier transform to
Eq.~\eqref{kernel} gives
\begin{equation}
\widetilde{\tau}(\mathbf{k})
=
\widetilde{H}(\mathbf{k})\widetilde{u}(\mathbf{k})
+
\widetilde{\eta}(\mathbf{k}),
\label{fourier-forward}
\end{equation}
where $\widetilde{H}$ is the Fourier transform of the
convolution kernel associated with $H$. Under the
discrete convolution and boundary convention used
for FFT-based evaluation, the convolution operator
is diagonalized by the Fourier transform. Conventional
regularized LCT inversion can therefore be written as
\begin{equation}
\min_{\widetilde{u}}
\left\{
\big\|
\widetilde{H}\odot\widetilde{u}
-\widetilde{\tau}
\big\|_2^2
+
\alpha\|\widetilde{u}\|_2^2
\right\},
\label{min-prob}
\end{equation}
where $\alpha>0$ is a frequency-independent regularization parameter and $\odot$ denotes pointwise multiplication. For an array $v$ on the Fourier grid, $\|v\|_2^2=\sum_{\mathbf{k}}|v(\mathbf{k})|^2$. The objective separates across frequency components and admits the solution
\begin{equation}
\widetilde{u}_{\mathrm{LCT}}^\star(\mathbf{k})
=
\frac{\overline{\widetilde{H}(\mathbf{k})}}
{|\widetilde{H}(\mathbf{k})|^2+\alpha}
\widetilde{\tau}(\mathbf{k}),
\label{sol}
\end{equation}
where the overline denotes complex conjugation. The analytical response and scalar regularization provide an efficient inverse filter. However, the analytical response may not fully represent practical measurements, and a single regularization strength
cannot independently adjust the treatment of different
frequency components. This can limit reconstruction quality when spectral reliability varies across frequencies or acquisition conditions.

\begin{figure}[t]
\begin{center}			
\includegraphics[width=0.9\linewidth]{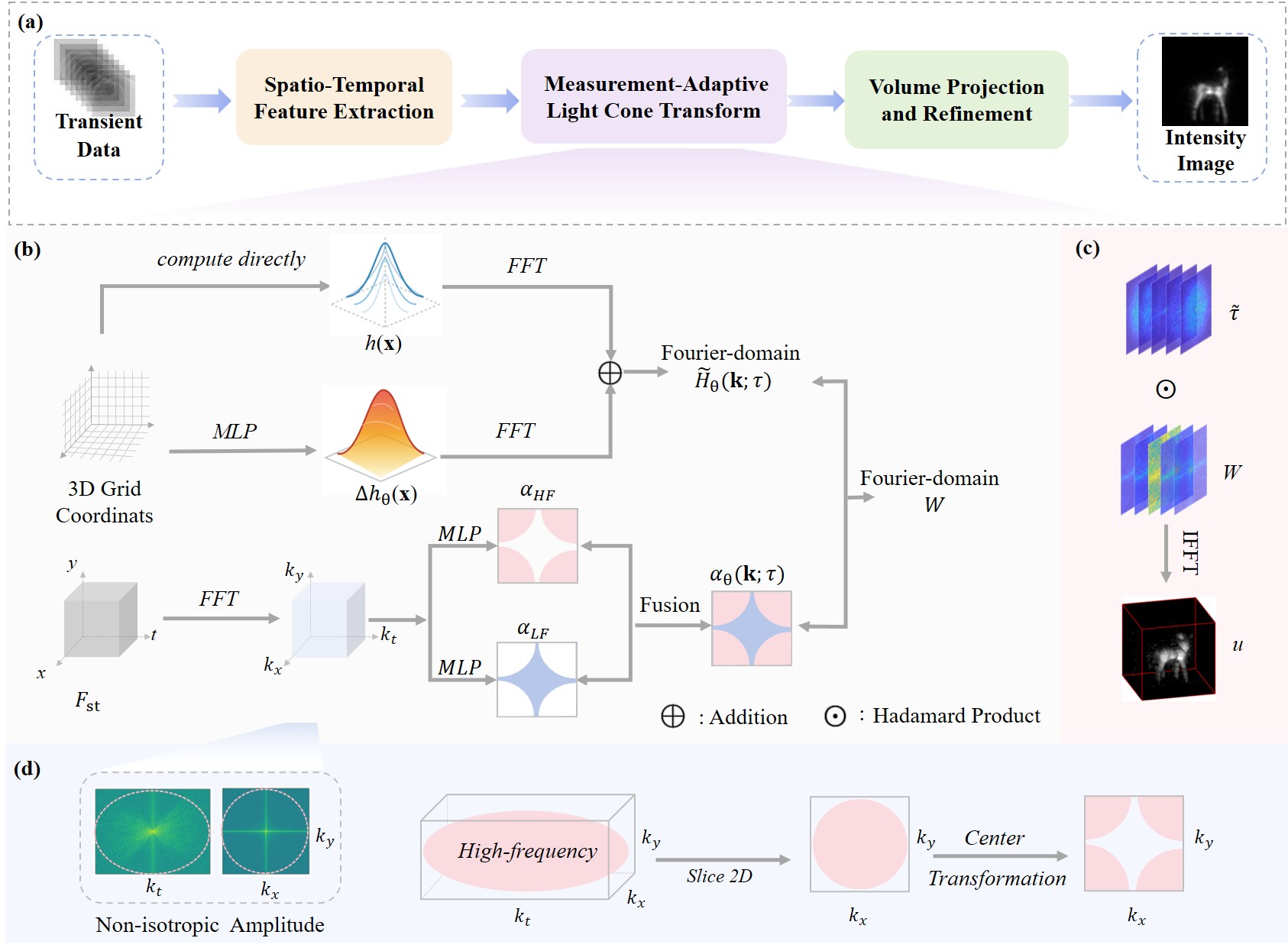}
\end{center}
\vspace{-0.5cm}
\caption{The architecture of our proposed MALCT framework.}
\label{network}
\end{figure}

{To introduce data-driven adaptation while retaining
pointwise Fourier-domain inversion, we combine the
analytical response with a shared learned response:
\begin{equation}
\widetilde{H}_{\Theta}(\mathbf{k};\tau)
=
\lambda_{\Theta}(\tau)\widetilde{H}(\mathbf{k})
+
[1-\lambda_{\Theta}(\tau)]
\Delta\widetilde{H}_{\Theta}(\mathbf{k}),
\label{learned-parameters}
\end{equation}
where $\Theta$ collectively denotes the trainable
parameters and $\lambda_{\Theta}(\tau)\in(0,1)$
is a measurement-conditioned mixing coefficient.
The learned branch
$\Delta\widetilde{H}_{\Theta}(\mathbf{k})$
is shared across inputs after training, whereas its
weight relative to the analytical response depends
on the input transient.  Equation~\eqref{learned-parameters} thus provides a restricted form of response adaptation: each measurement determines the mixing weight, but does not generate an unrestricted new kernel. For a fixed input, the effective response remains diagonal in the Fourier
domain and represents a shift-invariant convolution in the resampled domain.

We further replace the scalar $\alpha$ with a positive
regularization map $\alpha_{\Theta}(\mathbf{k};\tau)>0$, conditioned on both frequency and the input measurement. This map
allows different spectral components to receive different regularization strengths. For a given transient $\tau$, the predicted response and regularization map define
\begin{equation}
\min_{\widetilde{u}}
\left\{
\big\|
\widetilde{H}_{\Theta}(\cdot;\tau)
\odot\widetilde{u}
-\widetilde{\tau}
\big\|_2^2
+
\big\|
\alpha_{\Theta}^{1/2}(\cdot;\tau)
\odot\widetilde{u}
\big\|_2^2
\right\},
\label{min}
\end{equation}
where the dot denotes the full frequency grid.
The regularization term is equivalently
$\|
\alpha_{\Theta}^{1/2}(\cdot;\tau)
\odot\widetilde{u}
\|_2^2=
\sum_{\mathbf{k}}
\alpha_{\Theta}(\mathbf{k};\tau)
|\widetilde{u}(\mathbf{k})|^2$. 
For each input, the predicted response and regularization
map are held fixed during minimization with respect
to $\widetilde{u}$. Because
$\alpha_{\Theta}(\mathbf{k};\tau)>0$ at every frequency,
the quadratic objective has a unique minimizer:
\begin{equation}
\widetilde{u}_{\Theta}^\star(\mathbf{k})=
G_{\Theta}(\mathbf{k};\tau)
\widetilde{\tau}(\mathbf{k}) \quad \mbox{with}~~
G_{\Theta}(\mathbf{k};\tau) =
\frac{
\overline{\widetilde{H}_{\Theta}(\mathbf{k};\tau)}
}{
|\widetilde{H}_{\Theta}(\mathbf{k};\tau)|^2
+
\alpha_{\Theta}(\mathbf{k};\tau)
}.
\label{solution-MALCT}
\end{equation}
The effective inverse filter $G_{\Theta}$ is jointly
determined by response weighting and regularization.
The former balances the analytical and shared learned
responses, while the latter controls the penalty on
each spectral component, together enabling
measurement-adaptive reconstruction. For a fixed input,
inversion proceeds pointwise in the Fourier domain;
the overall mapping remains nonlinear because the
filter depends on the measurement. 
The effective response is optimized for reconstruction,
rather than explicit identification of individual
physical mismatch sources. Changes in the inverse
kernel reflect the combined effects of response
weighting and regularization and cannot be attributed
solely to correction of the physical propagation operator.} 

Finally, the reconstructed volume in the resampled
domain is
\[
u_{\Theta}^\star
=
\mathcal{F}^{-1}
\!\left[\widetilde{u}_{\Theta}^\star\right].
\]
The corresponding volume on the original depth grid
is obtained by applying $R_z^{-1}$.

\subsection{Our network architecture}\label{network-archi}

Based on the variational formulation in Eq.~\eqref{min},
we develop a reconstruction network consisting of three
modules: spatio-temporal feature extraction (STFE),
measurement-adaptive LCT inversion, and volume projection,
as illustrated in Figure~\ref{network}.
The network takes a resampled transient measurement
$\tau\in\mathbb{R}^{C\times T\times H\times W}$ as input,
where $C$ is the number of channels, $T$ is the number
of resampled temporal bins, and $H\times W$ denotes
the spatial sampling grid on the relay wall. The STFE module combines multiscale convolutional features with temporal and spatial attention to obtain a measurement representation. This representation conditions the response mixing coefficient and the
frequency-dependent regularization map. The MALCT inversion module then constructs the effective Fourier response by combining the analytical and shared learned responses and performs the inversion in Eq.~\eqref{solution-MALCT}. Finally, the volume projection module maps the reconstructed volume to two-dimensional intensity and depth estimates for supervision.
Together, these modules integrate learned measurement-dependent parameters with the analytical Fourier-domain inversion structure of LCT.

\begin{figure}[t]
      \begin{center}\includegraphics[width=0.7\linewidth]{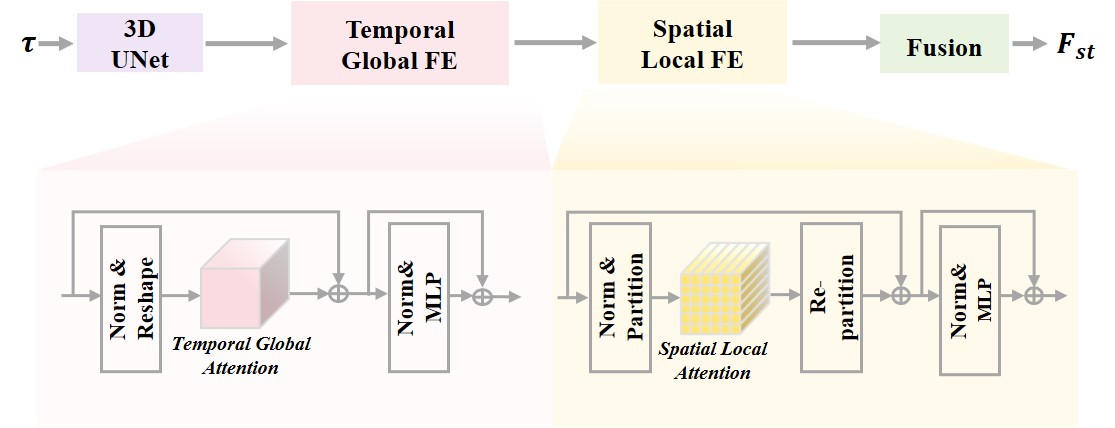}
	  \end{center}
      \vspace{-0.5cm}
	\caption{The spatio-temporal feature extraction module consists of a 3D UNet, temporal global feature extraction, spatial local feature extraction and fusion module. }
	\label{spatio-Temporal Feature Extraction}
\end{figure}

\subsubsection{spatio-temporal feature extraction}
As shown in Figure~\ref{spatio-Temporal Feature Extraction},
STFE first employs a three-dimensional convolutional
U-Net to extract multiscale features $f_c
=
\mathcal{C}_{\mathrm{UNet}}(\tau)
\in
\mathbb{R}^{4C\times\frac{T}{2}\times\frac{H}{2}
\times\frac{W}{2}}$. 
These features encode local temporal patterns and
spatial correlations in the transient measurements.
To complement convolutional feature extraction with
long-range temporal interactions, we apply global
temporal multi-head self-attention (G-TSA):
\[
f_{\mathrm G}
=
\mathcal{A}_{\mathrm{G\text{-}TSA}}(f_c).
\]
The resulting features are subsequently processed
by local spatial self-attention (L-SSA), which models
interactions within neighboring spatial regions:
\[
f
=
\mathcal{A}_{\mathrm{L\text{-}SSA}}(f_{\mathrm G}).
\]
Both modules preserve the input feature dimensions
and comprise normalization, attention, and MLP blocks
with residual connections. They differ in their token
organization and attention range: G-TSA aggregates
information across the temporal dimension, whereas
L-SSA performs attention within local spatial windows
and reassembles the resulting features. Finally, the convolutional features $f_c$ and attention-enhanced features $f$ are concatenated along the channel dimension and fused through a
$1\times1\times1$ convolution:
\[
\bm{F}_{\mathrm{st}}
=
\mathcal{C}_{1\times1\times1}
\!\left(\operatorname{Concat}(f_c,f)\right)
\in
\mathbb{R}^{4C\times\frac{T}{2}\times\frac{H}{2}
\times\frac{W}{2}}.
\]
The fused representation combines multiscale convolutional information with temporal and spatial attention features. It is used to predict the measurement-conditioned mixing coefficient
$\lambda_{\Theta}(\tau)$ and regularization map $\alpha_{\Theta}(\mathbf{k};\tau)$. The learned response branch itself is shared across measurements after training; its contribution to
the effective response varies through $\lambda_{\Theta}(\tau)$.

\subsubsection{Measurement-adaptive light-cone transform inversion}

The MALCT inversion module constructs the effective Fourier response $\widetilde{H}_{\Theta}(\mathbf{k};\tau)$
and predicts the regularization map $\alpha_{\Theta}(\mathbf{k};\tau)$ used in Eq.~\eqref{solution-MALCT}.
It retains the analytical LCT response as a reference and introduces learned response weighting and regularization to adapt the inverse filter to the input measurement. 

We first obtain a compact measurement representation
from the fused STFE features $z_{\tau} =\operatorname{GAP}(\bm{F}_{\mathrm{st}})$, where $\operatorname{GAP}$ denotes global average pooling over the temporal and spatial dimensions.
This representation provides the measurement-dependent input to the response-weighting and regularization networks. The shared learned response is parameterized by a coordinate MLP evaluated on the resampled grid and transformed into the Fourier domain:
\begin{equation}
\Delta\widetilde{H}_{\Theta}(\mathbf{k})
=
\mathcal{F}\!\left[
\operatorname{MLP}_H(\mathbf{x};\Theta_H)
\right](\mathbf{k}),
\label{shared-learned-response}
\end{equation}
where $\mathbf{x}$ denotes the three-dimensional resampled
coordinates. The coordinate MLP defines a response shared
across measurements after training, rather than an explicit
physical PSF estimate. Following Eq.~\eqref{learned-parameters},
this response is combined with the analytical response using
the measurement-conditioned coefficient
$\lambda_{\Theta}(\tau)=
\operatorname{sigmoid}\!\left(
\operatorname{MLP}_{\lambda}(z_{\tau};\Theta_{\lambda})
\right)\in(0,1)$. Thus, measurement dependence enters through the mixing weight, while the learned branch remains fixed during inference.

To account for variations in spectral reliability,
we additionally predict a frequency-dependent
regularization map:
\begin{equation}
\alpha_{\Theta}(\mathbf{k};\tau)
=
\operatorname{sigmoid}\!\left[
\operatorname{MLP}_{\mathrm{SNR}}
\!\left(
\gamma_{\mathrm{ell}}(\mathbf{k}),
\mathcal F(\mathbf F_{\mathrm{st}})(\mathbf k);\Theta_{\alpha}
\right)
\right]
+\epsilon,
\label{learned-adaptive-alpha}
\end{equation}
where $\epsilon>0$ provides a positive lower bound,
giving
$\alpha_{\Theta}(\mathbf{k};\tau)\in(\epsilon,1+\epsilon)$.
The ellipsoidal spectral encoding
$\gamma_{\mathrm{ell}}(\mathbf{k})$, illustrated in
Fig.~\ref{network}(d), accounts for the different
scales of the spatial and resampled axial frequency
coordinates. Together with $\mathbf F_{\mathrm{st}}$, this encoding
allows the network to assign regularization strengths
according to both frequency and measurement features.
The network $\operatorname{MLP}_{\mathrm{SNR}}$ predicts
regularization weights rather than explicit SNR estimates.

Finally, the effective response and regularization
map are substituted into Eq.~\eqref{solution-MALCT}
to form the measurement-conditioned inverse filter.
An inverse Fourier transform yields the reconstructed
volume in the resampled domain, and $R_z^{-1}$ maps
the result to the original depth grid.

\subsubsection{Volume projection and refinement} 
High-fidelity volumetric ground truth is difficult
to obtain for real NLOS scenes%
~\cite{chen2020learned,li2023nlost}.
We therefore project the reconstructed volume
$u(x,y,z)$ onto two-dimensional intensity and depth
maps for image-space refinement and supervision.
Let
$j^\star(x,y)=
\operatorname*{arg\,max}_{j\in\{1,\ldots,N_z\}}
u(x,y,z_j)$
denote the index of the maximum response along the
depth axis, where $z_j$ is the physical depth of
the $j$th sample. The initial projections are
\begin{equation}
\overline{I}(x,y)=
u\bigl(x,y,z_{j^\star(x,y)}\bigr),\qquad
\overline{D}(x,y)=z_{j^\star(x,y)},
\label{proj}
\end{equation}
where $\overline{I}$ records the maximum reconstructed intensity along each depth profile, and $\overline{D}$ gives the corresponding physical depth. These raw projections are refined employing $I$ and $D$ generated by ResNet-based modules.

\subsubsection{Physics-informed loss function}
{{We adopt a two-stage training strategy
(Algorithm~\ref{alg:two-stage-training}).
The first stage optimizes the STFE, MALCT inversion,
and intensity-refinement parameters, collectively
denoted by $\theta_I$. The second stage freezes
$\theta_I$ and optimizes the depth-refinement
parameters $\theta_D$.}} Let $u$ and $\tau$ denote the reconstructed volume and transient on the original depth and detector-time grids, respectively. The first-stage loss is
\begin{equation}
\begin{aligned}
\mathcal{L}_I(\theta_I)
=
\underbrace{
\left\|H_{\Theta}(\tau)R_z u-R_t\tau\right\|_2^2
}_{\mathcal{L}_{\mathrm{phy}}}
+
\underbrace{
\left\|I-I_{\mathrm{GT}}\right\|_2^2
}_{\mathcal{L}_{\mathrm{alb}}}
+
\gamma\mathcal{L}_{\mathrm{TV}}(u)
+
\mu\mathcal{L}_{\mathrm{Tik}}(\alpha_{\Theta}),
\end{aligned}
\label{eq:reconstruction_loss}
\end{equation}
where $\gamma,\mu\geq0$ control the regularization
weights. The term $\mathcal{L}_{\mathrm{phy}}$
encourages consistency with the measurement in the
resampled domain under the learned effective operator
$H_{\Theta}(\tau)$, while $\mathcal{L}_{\mathrm{alb}}$
supervises the refined intensity prediction $I$.
The TV penalty promotes local regularity in $u$,
and the Tikhonov penalty regularizes the predicted
map $\alpha_{\Theta}$.

In the second stage, the depth-refinement module
is trained using
\begin{equation}
\mathcal{L}_D(\theta_D)
=
\left\|D-D_{\mathrm{GT}}\right\|_2^2,
\label{eq:depth_loss}
\end{equation}
where $D$ is obtained by refining the initial depth projection $\overline{D}$ in Eq.~\eqref{proj}. Since the reconstruction network is frozen, depth-loss gradients update only the refinement module and do not propagate through the hard
$\arg\max$ operation.

\begin{algorithm}[h]
\caption{Two-stage training}
\label{alg:two-stage-training}
\begin{algorithmic}[1]
\State \textbf{Input}: Training data; iteration counts $K_I,K_D$
\State \textbf{Stage 1}: Train $\theta_I$; freeze $\theta_D$
\For{$k=1,\ldots,K_I$}
    \State Reconstruct $u$ via Eq.~\eqref{solution-MALCT}
    and inverse transforms
    \State Project via Eq.~\eqref{proj}; refine to obtain $I$
    \State Update $\theta_I$ via Eq.~\eqref{eq:reconstruction_loss}
\EndFor
\State \textbf{Stage 2}: Freeze $\theta_I$; train $\theta_D$
\For{$k=1,\ldots,K_D$}
    \State Reconstruct $u$ with frozen $\theta_I$
    \State Project via Eq.~\eqref{proj}; refine to obtain $D$
    \State Update $\theta_D$ via Eq.~\eqref{eq:depth_loss}
\EndFor
\State \textbf{Output}: Trained $\theta_I,\theta_D$
\end{algorithmic}
\end{algorithm}

\section{Numerical experiments}\label{result}
\subsection{Experimental setup}

\textbf{Datasets.}
We generate synthetic transient measurements using the physically based NLOS renderer introduced in \cite{chen2020learned}. The dataset is constructed from 277 motorcycle models in ShapeNetCore~\cite{chang2015shapenet}.
The simulated acquisition uses a $128\times128$ spatial sampling grid and 512 temporal bins, with a temporal bin width of $\Delta t=33$~ps. The resulting dataset contains 2,712 training samples
and 291 test samples. {{We also evaluate MALCT on two public real-world NLOS datasets. The Stanford dataset~\cite{lindell2019wave} provides measurements of the same scenes acquired over 180 and 10 min per full scan, enabling controlled acquisition-time comparisons. The USTC dataset~\cite{li2023nlost}, acquired at 8 ms per scanning point, assesses performance under a different system and acquisition configuration. }}

\textbf{Baseline selection.} We compare MALCT with six representative reconstruction methods: three physics-based methods, namely LCT ~\cite{o2018confocal}, f-k migration
(FK)~\cite{lindell2019wave}, and phasor field (PF)~\cite{liu2020phasor}; and three learning-based methods, namely LFE~\cite{chen2020learned}, NLOST~\cite{li2023nlost}, and
LPP~\cite{sun2025generalizable}. For a consistent training comparison, LFE, NLOST, and LPP are retrained from scratch using the same synthetic dataset and noise-generation settings as MALCT.

\begin{figure}[t]
      \begin{center}		\includegraphics[width=1\linewidth]{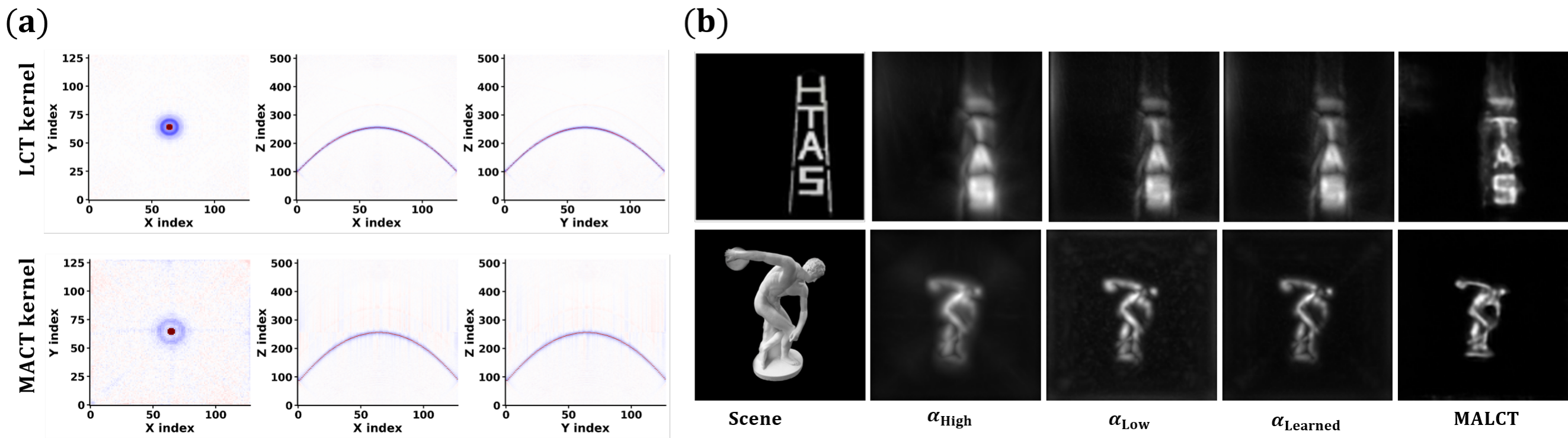}
	  \end{center}
      \vspace{-0.5cm}
	\caption{{{Results for the USTC letter and Stanford Statue scenes, with the Statue acquired over 10 min. (a) Central slices of the jointly normalized inverse kernels learned from the USTC letter scene. (b) \(64\times64\) albedo reconstructions using LCT with \(\alpha_{\mathrm{High}}=10\) or \(\alpha_{\mathrm{Low}}=0.1\) and MALCT with \(\alpha_{\mathrm{Learned}}\).}}}
\label{LCT-based}
\end{figure}

\textbf{Implementation details.}
Our MALCT is implemented in PyTorch and trained on a workstation equipped with two NVIDIA GeForce RTX 3090 GPUs. We use the AdamW optimizer with a batch size of 1 for 50 epochs: 40 epochs for
reconstruction and intensity prediction, followed by 10 epochs for depth refinement. A cosine annealing schedule decreases the learning rate from $1\times10^{-4}$ to $1.5\times10^{-5}$
over training. The loss weights are set to $\gamma=\mu=1\times10^{-7}$.

\subsection{Evaluation on inverse filter}
We first examine the effective inverse kernels and
the influence of regularization on reconstruction.
Figure~\ref{LCT-based}(a) compares central slices of
the LCT and MALCT inverse kernels. MALCT retains
the overall light-cone-like structure while exhibiting
localized differences near the center and cone surface.
These differences reflect the combined effects of
the effective response and regularization; the kernel
visualization alone does not establish correction
of physical model mismatch.

Figure~\ref{LCT-based}(b) illustrates the associated
reconstruction trade-off. With weak regularization,
LCT can amplify unreliable spectral components,
whereas strong regularization can suppress weak
structures and fine details. Through
measurement-conditioned response weighting and
frequency-dependent regularization, MALCT produces
fewer visible artifacts and better-preserved structures
in the examples shown.

\begin{figure}[t]
      \begin{center}			
\includegraphics[width=1\linewidth]{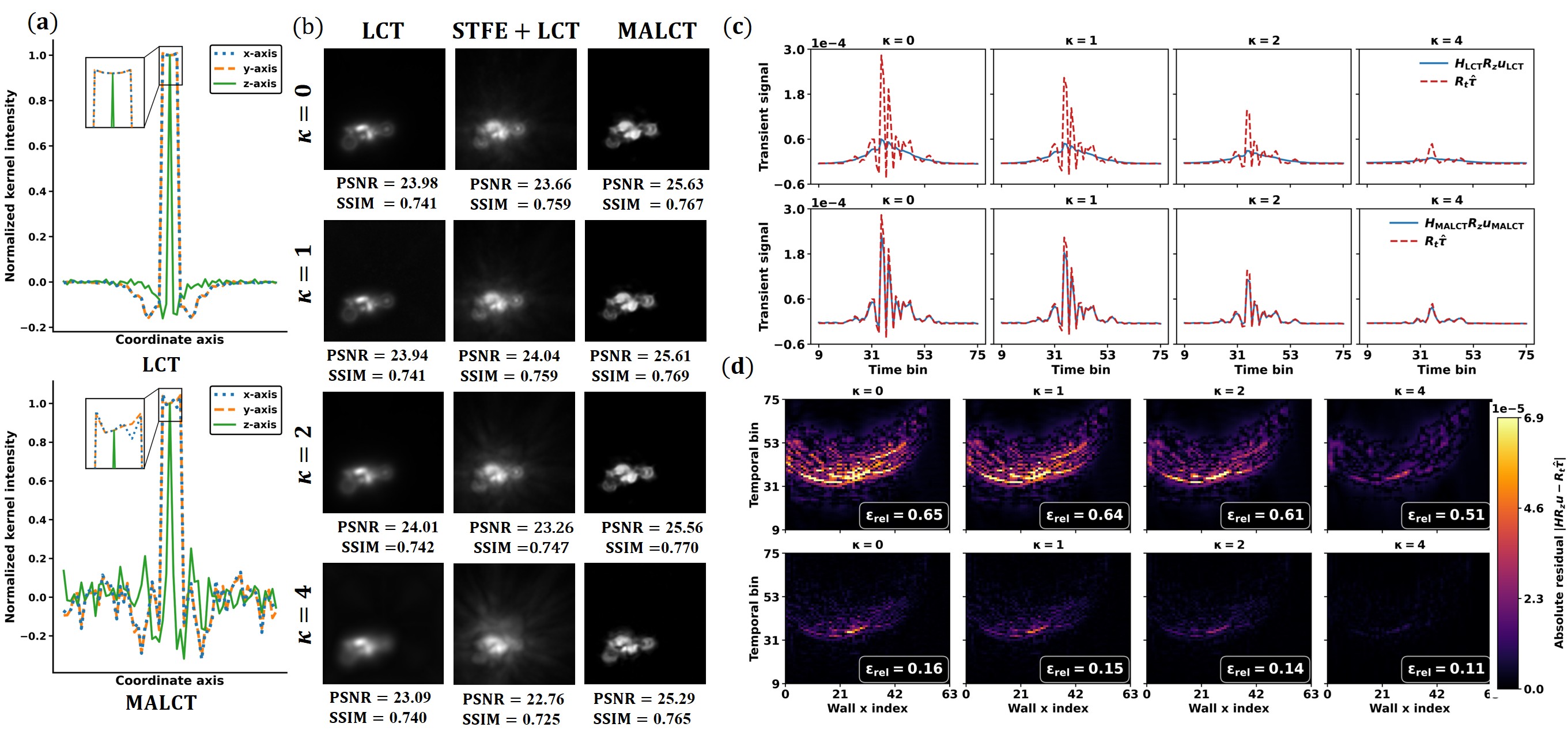}  \end{center}
\vspace{-0.5cm}
\caption{{Evaluation under prescribed Gaussian IRF broadening. (a)1D inverse kernel profiles ($\kappa=1$) for LCT and MALCT.
(b) Intensity reconstructions with PSNR/SSIM values.
(c) Resampled STFE transients and corresponding reconstruction-projections.
(d) Absolute $t$-$x$ residuals and relative errors.}}
\label{fig:kernelIRF}
\end{figure}

\begin{table}[t]
\centering
\caption{
Quantitative comparison under different noise levels. PSNR and SSIM are computed for reconstructed intensity images, while RMSE and MAD are computed for depth maps. The best and second-best results are shown in \textbf{bold} and \underline{underline}, respectively.
}
\label{tab:noise}
\footnotesize
\renewcommand{\arraystretch}{1.0}
\setlength{\tabcolsep}{3.2pt}

\begin{tabular}{l|cccc|cccc|cccc}
\toprule
\multirow{3}{*}{\textbf{Method}}
& \multicolumn{4}{c|}{\textbf{SNR = 10 dB}}
& \multicolumn{4}{c|}{\textbf{SNR = 8 dB}}
& \multicolumn{4}{c}{\textbf{SNR = 5 dB}}
\\

\cmidrule(lr){2-5}
\cmidrule(lr){6-9}
\cmidrule(lr){10-13}

& \multicolumn{2}{c}{\textbf{Intensity}}
& \multicolumn{2}{c|}{\textbf{Depth}}
& \multicolumn{2}{c}{\textbf{Intensity}}
& \multicolumn{2}{c|}{\textbf{Depth}}
& \multicolumn{2}{c}{\textbf{Intensity}}
& \multicolumn{2}{c}{\textbf{Depth}}
\\

\cmidrule(lr){2-3}
\cmidrule(lr){4-5}
\cmidrule(lr){6-7}
\cmidrule(lr){8-9}
\cmidrule(lr){10-11}
\cmidrule(lr){12-13}

& \textbf{PSNR}$\uparrow$
& \textbf{SSIM}$\uparrow$
& \textbf{RMSE}$\downarrow$
& \textbf{MAD}$\downarrow$

& \textbf{PSNR}$\uparrow$
& \textbf{SSIM}$\uparrow$
& \textbf{RMSE}$\downarrow$
& \textbf{MAD}$\downarrow$

& \textbf{PSNR}$\uparrow$
& \textbf{SSIM}$\uparrow$
& \textbf{RMSE}$\downarrow$
& \textbf{MAD}$\downarrow$
\\

\midrule

LCT
& 21.672 & 0.480 & 0.215 & 0.168
& 21.569 & 0.480 & 0.215 & 0.168
& 21.267 & 0.471 & 0.215 & 0.168
\\

FK
& 26.452 & 0.677 & 0.112 & 0.059
& 26.414 & 0.726 & 0.112 & 0.059
& 26.268 & 0.617 & 0.111 & 0.059
\\

PF
& 26.765 & 0.367 & 0.112 & 0.027
& 25.998 & 0.279 & 0.112 & 0.027
& 25.689 & 0.341 & 0.133 & 0.064
\\

LFE
& 27.052 & 0.792 & 0.066 & 0.021
& 26.235 & 0.787 & 0.065 & \underline{0.021}
& 26.176 & \underline{0.789} & 0.065 & 0.021
\\

NLOST
& 27.867 & 0.881 & \underline{0.059} & 0.021
& 27.768 & 0.834 & \underline{0.061} & 0.023
& 26.971 & 0.739 & 0.086 & 0.034
\\

LPP
& \underline{28.183} & \underline{0.888} & 0.060 & \underline{0.019}
& \underline{28.011} & 0.869 & 0.065 & 0.022
& \underline{27.008} & 0.766 & \underline{0.063} & \underline{0.020}
\\

\midrule

\textbf{MALCT}
& \textbf{28.583} & \textbf{0.916} & \textbf{0.058} & \textbf{0.016}
& \textbf{28.580} & \textbf{0.912} & \textbf{0.058} & \textbf{0.016}
& \textbf{28.540} & \textbf{0.909} & \textbf{0.059} & \textbf{0.016}
\\

\bottomrule
\end{tabular}
\end{table}

{{To complement these observations with a controlled
evaluation, we introduce temporal-response mismatch
by convolving clean transients with normalized
Gaussian IRFs before LCT temporal resampling:
\[
\tau_\kappa(\mathbf{x},t)
=
\bigl(g_\sigma *_t\tau\bigr)(\mathbf{x},t)
=
\int_{-\infty}^{\infty}
g_\sigma(t-t')\tau(\mathbf{x},t')\,\mathrm{d}t',
\]
where $\tau$ is the clean transient  and
$\sigma=\kappa\Delta t$ with $\kappa\in\{0,1,2,4\}$
and $\Delta t$ being the temporal bin width.
For $\sigma>0$, the Gaussian IRF is $g_\sigma(t)
=
\frac{1}{\sqrt{2\pi}\sigma}
\exp\!\left(-\frac{t^2}{2\sigma^2}\right)$. 
For $\kappa=0$, we set $g_0(t)=\delta(t)$, yielding the unperturbed transient $\tau_0=\tau$.  Figure~\ref{fig:kernelIRF}(a) shows representative inverse-kernel profiles at $\kappa=1$. Both methods exhibit a central peak and negative sidelobes,
but differ in their profile shapes. Figure~\ref{fig:kernelIRF}(b) compares reconstruction quality across the prescribed IRF widths. Our MALCT reduces visible blurring and achieves higher PSNR and SSIM than the compared methods at all tested perturbation levels.

As a complementary diagnostic, we evaluate
reconstruction-projection consistency for STFE+LCT
and MALCT using the same processed transient
$\widehat{\tau}_\kappa=\operatorname{STFE}(\tau_\kappa)$.
For each method and perturbation level, we compute
\[
u_\kappa
=R_z
\mathcal{F}^{-1}\!\left[
G_\kappa\odot\mathcal{F}(R_t\widehat{\tau}_\kappa)
\right],
\qquad
\varepsilon_{\mathrm{rel}}
=
\frac{
\|H_\kappa u_\kappa-R_t\widehat{\tau}_\kappa\|_2
}{
\|R_t\widehat{\tau}_\kappa\|_2
},
\]
where $u_\kappa$ denotes the reconstructed volume in the resampled domain, and $H_\kappa$ and $G_\kappa$ are the corresponding method's forward convolution operator and Fourier-domain inverse filter. The dependence of $\varepsilon_{\mathrm{rel}}$ on $\kappa$ is omitted for brevity.
Holding the STFE output fixed controls for differences in feature extraction, while the inversion modules can differ in both response and regularization. Figures~\ref{fig:kernelIRF}(c) and \ref{fig:kernelIRF}(d) show that MALCT reprojections
more closely match the resampled STFE output and yield lower relative residuals. These results complement the reconstruction-quality comparison under controlled IRF broadening, with the residuals measuring consistency under each method's respective forward model.
}}

\begin{figure}[t]
    \begin{center}			
     \includegraphics[width=0.85\linewidth]{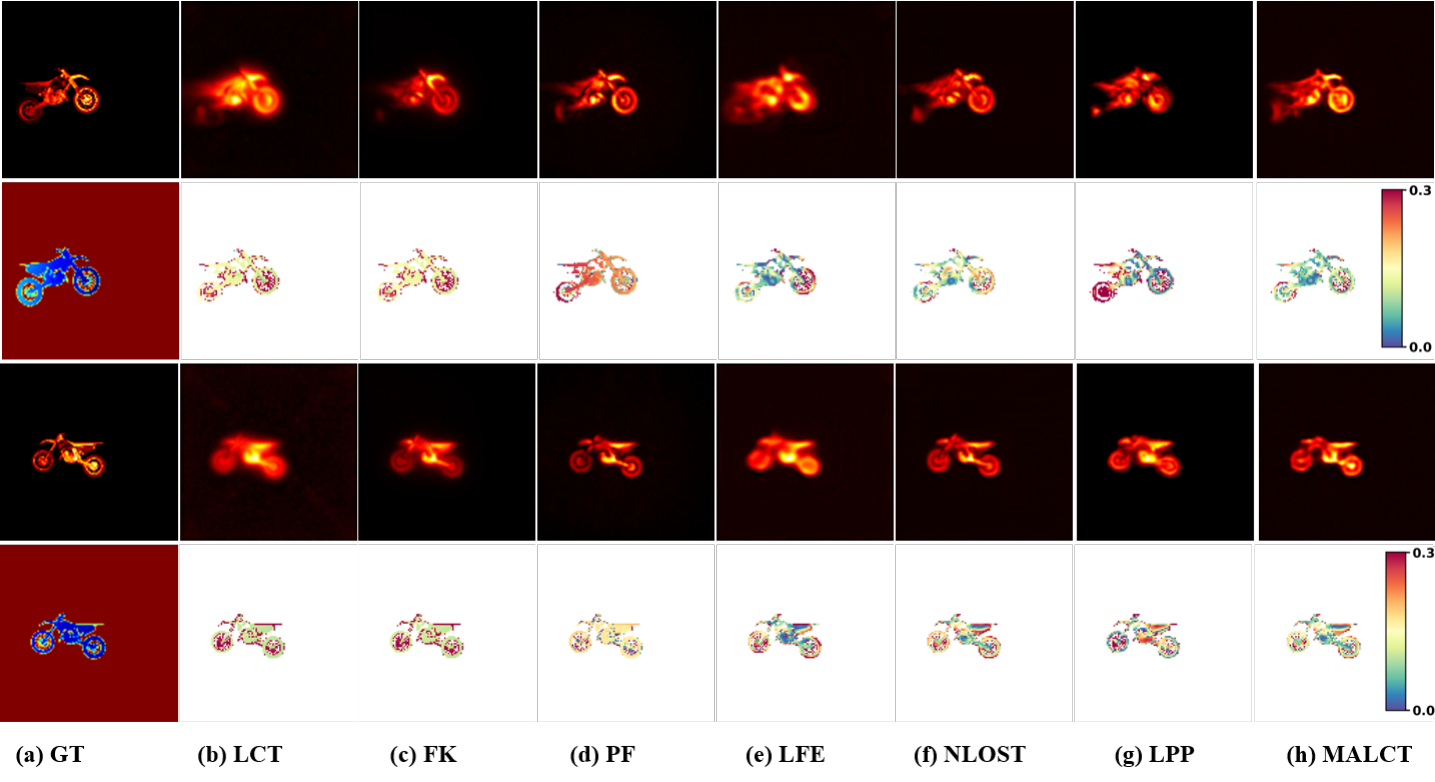}
	  \end{center}
      \vspace{-0.5cm}
	\caption{Qualitative comparison of the reconstructed intensity images and the depth error maps, where the depth errors are visualized in the range $[0, 0.3]$, within the ROI on the simulated dataset.}
	\label{simulate}
\end{figure}

\begin{table}[t]
\centering
\caption{
Computational efficiency comparison in terms of inference time and GPU memory usage. The best and second-best results are shown in \textbf{bold} and \underline{underline}, respectively.
}
\label{tab:efficiency}
\footnotesize
\renewcommand{\arraystretch}{1.0}
\setlength{\tabcolsep}{7pt}

\begin{tabular}{l|ccccccc}
\toprule
\textbf{Metric}
& \textbf{LCT}
& \textbf{FK}
& \textbf{PF}
& \textbf{LFE}
& \textbf{NLOST}
& \textbf{LPP}
& \textbf{MALCT}
\\
\midrule
\textbf{Time (s)}$\downarrow$
& \underline{0.05}
& 0.06
& 0.09
& \textbf{0.04}
& 0.42
& 0.25
& 0.12\\
\textbf{Memory (GB)}$\downarrow$
& 8.2
& 9.7
& 11.6
& \textbf{3.3}
& 21.4
& 9.1
& \underline{5.5}\\
\bottomrule
\end{tabular}
\end{table}

\subsection{Experimental results on synthetic data}
{ We generate noisy transients by adding a constant
background offset and additive Gaussian noise,
followed by a multiplicative perturbation:
\[
\tau_{\mathrm g}
=
\tau+b\mathbf{1}+c_1\boldsymbol{\epsilon}_1,
\qquad
\tau_{\mathrm{noisy}}
=
\tau_{\mathrm g}
+
c_2(\tau_{\mathrm g}\odot\boldsymbol{\epsilon}_2),
\]
where $\tau$ is the clean transient,
$\boldsymbol{\epsilon}_1$ and $\boldsymbol{\epsilon}_2$
are independent standard Gaussian random vectors,
and $\odot$ denotes elementwise multiplication.
The parameters $b$, $c_1$, and $c_2$ control the
background offset, additive-noise standard deviation,
and multiplicative-noise strength, respectively.
The baseline configuration uses $b=c_1=c_2=0.02$.
Additional controlled tests separately increase
$b$ or $c_1$ to $0.04$, keeping the remaining
parameters fixed. These tests assess sensitivity
to individual parameter changes, which also affect
the variance of the subsequent multiplicative
perturbation.}

\begin{figure}
    \centering
    \includegraphics[width=0.85\linewidth]{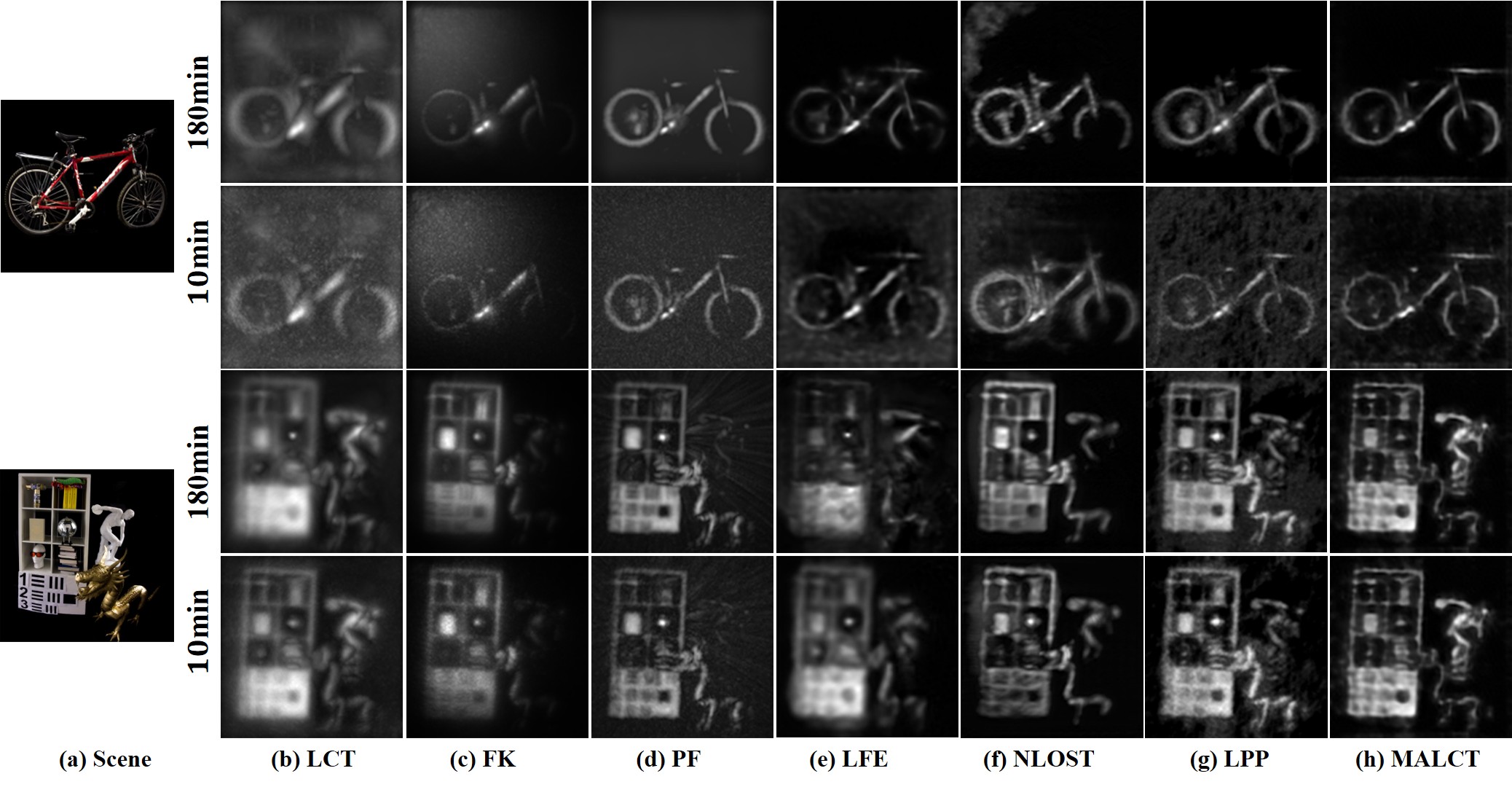}
    \vspace{-0.2cm}
    \caption{{Reconstructions from the Stanford dataset under 180-minute and 10-minute exposures.}}
    \label{fig:placeholder}
\end{figure}

\begin{figure}[t]
      \begin{center}			
\includegraphics[width=0.8\linewidth]{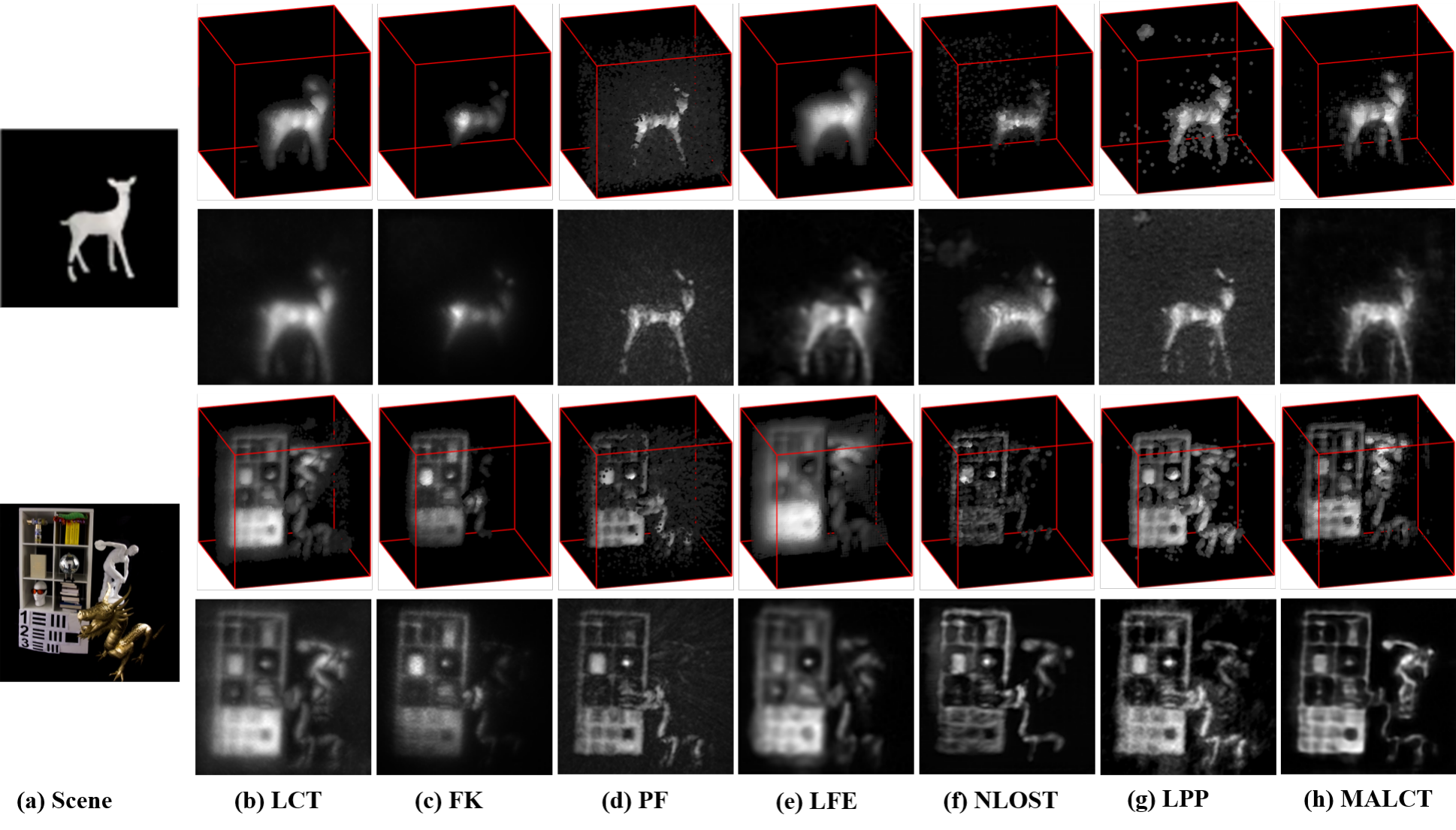}
	  \end{center}
      \vspace{-0.5cm}
	\caption{Reconstructed hidden scenes and corresponding 2D albedo images from the USTC and Stanford 10-minute  real-world datasets.}
	\label{fullsample}
\end{figure}

Table~\ref{tab:noise} summarizes reconstruction performance
on synthetic transients at different noise levels.
MALCT achieves better intensity and depth metrics than
the physics-based baselines and the learning-based
methods LFE, NLOST, and LPP under the tested conditions.
Its PSNR advantage over LPP increases from $0.40$~dB
at an SNR of $10$~dB to $1.53$~dB at $5$~dB,
supporting the noise robustness of the complete
reconstruction framework. Table~\ref{tab:efficiency} reports inference time and GPU memory usage. MALCT requires more computation and memory than LFE, but less than NLOST and LPP,
providing a balance between reconstruction quality and computational cost.

\begin{figure}[t]
      \begin{center}			
\includegraphics[width=0.8\linewidth]{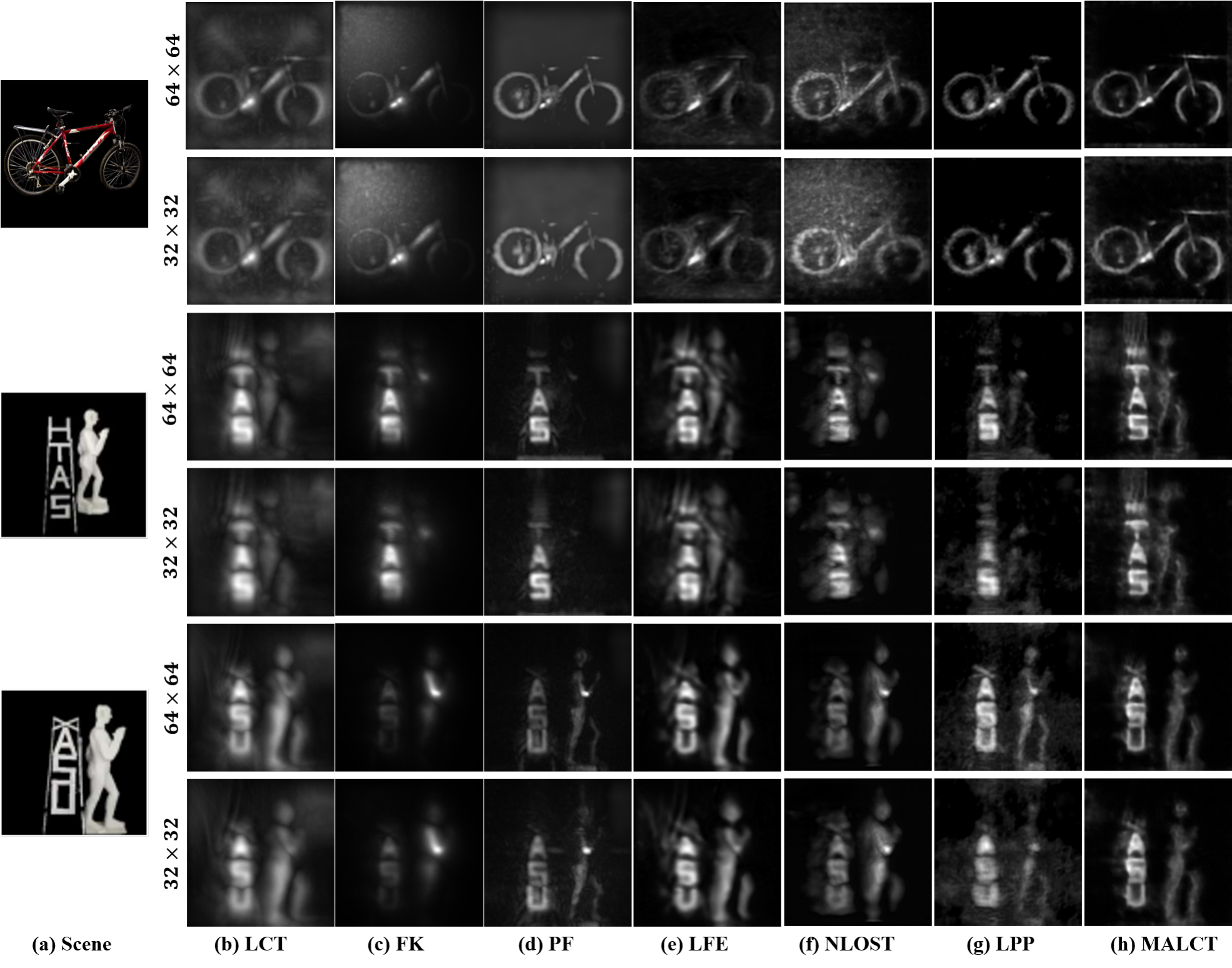}
	  \end{center}
      \vspace{-0.5cm}
	\caption{{Reconstructed 2D albedo images from the Stanford 180-minute and USTC datasets under uniform spatial subsampling. Measurements are resized to \(128\times128\) by nearest-neighbor interpolation and processed by the same model without retraining.}}
	\label{lowsample}
\end{figure}

Figure~\ref{simulate} compares reconstructed intensity
images and depth error maps. In the examples shown,
the LCT, FK, and PF exhibit blurred or missing structures
and larger depth errors. The LFE, NLOST, and LPP improve
structural clarity but retain local artifacts and
depth inaccuracies. Our MALCT better preserves fine
structures and reduces depth errors, consistent
with the quantitative results in
Table~\ref{tab:noise}.

\subsection{Experimental results on real-world  data}

{{Figures~\ref{fig:placeholder}, \ref{fullsample} and
\ref{lowsample} evaluate reconstruction across acquisition
times, imaging systems, and spatial sampling densities.
The Stanford measurements acquired over 180 and 10 min
provide a same-scene comparison under reduced acquisition
time. The USTC dataset evaluates transfer to a different
acquisition system, while spatially downsampled measurements assess performance with fewer scanning points.
}} 

In the examples shown, MALCT preserves more complete
object structures with fewer visible artifacts.
LCT and LFE produce blurred reconstructions, while
FK recovers the main structures but loses finer features.
NLOST improves structural clarity but leaves some
details unresolved. The PF and LPP retain local details
alongside residual artifacts. Our MALCT provides a favorable
balance between structural completeness and artifact
suppression across the evaluated conditions, including
shorter acquisition times and spatial sampling grids
down to $32\times32$. These qualitative results support
its applicability to limited measurements and different
acquisition systems.

\begin{figure*}[t]
\centering

\begin{subfigure}{0.75\textwidth}
\centering
\phantomsubcaption
\label{fig:ablation_table}
\makebox[\linewidth][l]{\textbf{(a)}}\par
\vspace{2pt}

\small
\renewcommand{\arraystretch}{1.0}
\setlength{\tabcolsep}{3pt}
\begin{tabular*}{\linewidth}
{@{\extracolsep{\fill}}ccccc@{}}
\hline
\textbf{Variant}
& \textbf{Attention}
& \textbf{Learned PSF}
& \textbf{Globally learned $\boldsymbol{\alpha}$}
& \textbf{Frequency-adaptive $\boldsymbol{\alpha}$} \\
\hline
I
& $\times$ & $\times$ & $\times$ & $\times$ \\
II
& $\checkmark$ & $\times$ & $\times$ & $\times$ \\
III
& $\times$ & $\checkmark$ & $\times$ & $\times$ \\
IV
& $\checkmark$ & $\checkmark$ & $\times$ & $\times$ \\
\hline
V
& $\checkmark$ & $\checkmark$ & $\checkmark$ & $\times$ \\
VI
& $\checkmark$ & $\checkmark$ & $\times$ & $\checkmark$ \\
\hline
\end{tabular*}
\end{subfigure}

\par\vspace{4pt}

\begin{subfigure}{0.85\textwidth}
\centering
\phantomsubcaption
\label{fig:ablation_results}
\makebox[\linewidth][l]{\textbf{(b)}}\par
\vspace{2pt}

\includegraphics[width=\linewidth]{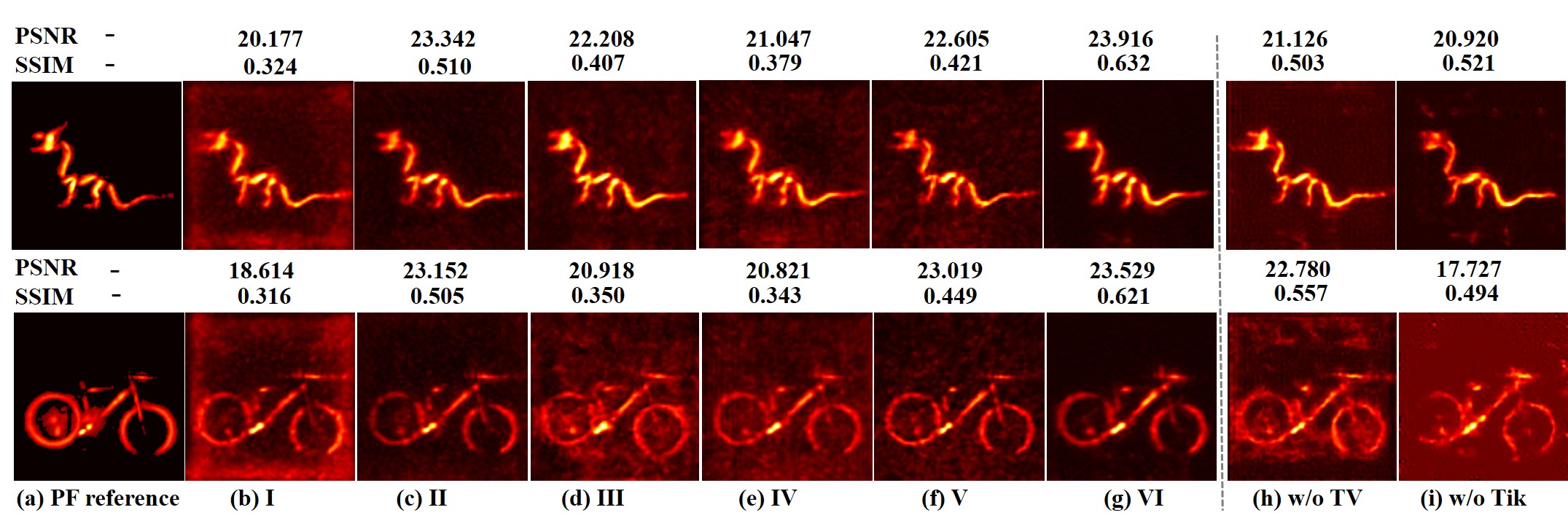}
\end{subfigure}

\caption{
{Ablation study on Dragon and Bike.
(a) Configurations. All variants retain the 3D convolutional
backbone; NoAtt removes G-TSA and L-SSA. Variants I-IV
use fixed scalar regularization, V a globally learned
scalar, and VI a measurement-conditioned frequency map.
(b) Reconstructions for Variants I-VI and the model without \(\mathcal L_{\mathrm{TV}}\) or \(\mathcal L_{\mathrm{Tik}}\) from $32\times32$ measurements.
PSNR and SSIM are computed against the PF reference
obtained from the original $512\times128\times128$, 180-min
measurements.}
}
\label{ablation}
\end{figure*}

{{Figure~\ref{ablation} examines the contributions of
attention, the learned response and regularization.
All variants retain the 3D convolutional backbone.
Variant I uses the analytical LCT response with fixed
scalar regularization and no attention. Variants II
and III add attention and the learned response,
respectively, while Variant IV combines both.
Variants V and VI retain both components and replace
fixed regularization with a globally learned scalar
and a measurement-conditioned frequency-dependent
map, respectively. Compared with Variant I, Variant II shows improved
structural continuity and fewer background artifacts,
supporting the contribution of temporal and spatial
attention. Introducing the learned response alters
local details, but does not consistently improve PSNR
or SSIM under fixed scalar regularization. In particular,
the degradation from Variant II to IV indicates that
response learning alone is insufficient to improve
overall reconstruction quality. Learning a global
regularization scalar in Variant V partially recovers
this loss, while Variant VI achieves the highest PSNR
and SSIM for both scenes and better preserves their
structures. These results support the complementary
roles of response learning and adaptive regularization.
In the loss ablations, removing \(\mathcal L_{\mathrm{TV}}\) increases background artifacts, consistent with its role in constraining spatial variation of the reconstructed volume. Removing \(\mathcal L_{\mathrm{Tik}}\) weakens the recovered structures, particularly for Bike. The metrics quantify agreement with the high-resolution PF reference.}}

\section{Conclusion}\label{conclusion}
We presented MALCT, a physics-guided framework that
retains the Fourier-domain structure of LCT while
adapting its inverse filter through measurement-conditioned
response weighting and frequency-dependent regularization.
Experiments on synthetic and real measurements show
improved reconstruction quality with competitive
computational cost. Controlled evaluations further
support robustness to IRF broadening, reduced acquisition
time and sparse spatial sampling. 

Future work will investigate extensions to non-confocal geometries and multi-bounce transport, where the standard LCT assumptions may no longer hold. We will also investigate joint or blind estimation of the temporal instrument response function to further improve robustness to detector jitter, timing uncertainty, and hardware synchronization errors. As shown in Fig.~\ref{fullsample}, while wave-aware formulations improve intensity recovery by modeling coherent propagation effects, they impose stricter demands on noise robustness, highlighting the need for robust reconstruction strategies in future work.

\begin{backmatter}
\bmsection{Acknowledgment} The work was supported by the National Natural Science Foundation of China (NSFC 12671658 and T2541053).

\bmsection{Disclosures}
The authors declare no conflicts of interest.

\bmsection{Data availability} Data underlying the results presented in this paper are available in Ref. \cite{chang2015shapenet}, 
Ref. \cite{lindell2019wave} and Ref. \cite{li2023nlost}.

\end{backmatter}

\bibliography{sample}






\end{document}